%% file: main.tex
\documentclass{article}
\usepackage{iclr2027_conference,times}
\usepackage[utf8]{inputenc}
\usepackage[T1]{fontenc}
\usepackage[hidelinks]{hyperref}
\usepackage{url,booktabs}
\usepackage{amsmath,amssymb,amsthm}
\usepackage{graphicx,tabularx,array,caption,placeins,xurl}
\usepackage{microtype}
\newcolumntype{Y}{>{\raggedright\arraybackslash}X}
\numberwithin{equation}{section}
\theoremstyle{definition}
\newtheorem{definition}{Definition}
\theoremstyle{plain}
\newtheorem{theorem}{Theorem}
\title{Control Foundation Models across Changing Block Diagrams: Generalizing Time-Series Prediction to Dynamic Graphs}
\author{%
  Yuji Okamoto\thanks{Corresponding author.}\\
  RIKEN\\
  2-1 Hirosawa, 351-0198, Japan\\
  \texttt{yuji.0001@gmail.com} \\
  \And
  Kenta Hoshino\\
  Institute of Science Tokyo,\\
  2-12-1 Ookayama, Meguro-ku,\\
  Tokyo 152-8550, Japan\\
  \texttt{hoshino@comp.isct.ac.jp} \\
  \And
  Masaki Inoue\\
  Keio University,\\
  3-14-1 Hiyoshi, Kohoku-ku,\\
  Yokohama, Kanagawa 223-8522, Japan\\
  \texttt{minoue@appi.keio.ac.jp} \\
}

\iclrfinalcopy
\makeatletter
\def\fps@figure{t}
\def\fps@table{t}
\makeatother
\begin{document}
\raggedbottom
\maketitle
\lhead{Preprint}
\input{sections/abstract}
\input{sections/introduction}
\input{sections/related-works}
\input{sections/background}
\input{sections/method}
\input{sections/results}
\FloatBarrier
\input{sections/conclusion}
\FloatBarrier
\section*{Acknowledgments}
This work was supported by JST Moonshot R\&D (JPMJMS2021), JST Research and Development Program for Next-generation Edge AI Semiconductors (JPMJES2511), JSPS KAKENHI (JP25K00148, JP25K01254, JP25H02626, JP26K14994, JP23K03902), a project (JPNP14004) commissioned by the New Energy and Industrial Technology Development Organization (NEDO), and RIKEN TRIP initiative (AGIS).

\bibliographystyle{iclr2027_conference}
{\small\sloppy\setlength{\bibsep}{2pt}\bibliography{references}}
\clearpage
\appendix
\setcounter{table}{0}
\setcounter{figure}{0}
\renewcommand{\thetable}{A\arabic{table}}
\renewcommand{\thefigure}{A\arabic{figure}}
\renewcommand{\theHfigure}{appendix.A.\arabic{figure}}
\renewcommand{\theHtable}{appendix.A.\arabic{table}}
\input{sections/appendix}
\FloatBarrier
\setcounter{section}{2}
\setcounter{figure}{0}
\setcounter{table}{0}
\renewcommand{\thefigure}{C\arabic{figure}}
\renewcommand{\thetable}{C\arabic{table}}
\renewcommand{\theHfigure}{appendix.C.\arabic{figure}}
\renewcommand{\theHtable}{appendix.C.\arabic{table}}
\input{sections/appendix-c}
\end{document}

%% file: sections/abstract.tex
\begin{abstract}

In the prediction and control of physical systems, combining records from different experiments and operations is expected to supply information that no single experiment provides on its own. Time-series foundation models, which learn a shared predictor from diverse time series, offer a framework for treating such records as a common training resource. In addition, advances in query-based prediction models, which switch between known signals and target signals within the same time series, have made it possible to handle response prediction and control-input generation in a unified manner. However, in control-experiment data, the connections between signals change with the experimental stage, such as system identification, filtering, and feedback control. Such data require a formulation that handles signal values and connection structures dynamically and in a unified way. In this study, we formulate a control foundation model. It defines a data space of dynamic graphs that contain connections, signal values, and missing values, together with an operation on this space that specifies the known signals and the prediction targets. This allows response prediction and control-input generation to be described as a common masked prediction problem, even for experimental data with different connections and recorded signals. The Block Diagram Transformer (BDT), which implements this formulation, combines attention along the time axis with a graph neural network that follows the block-diagram connections. It can use the pretrained parameters of existing time-series foundation models as initial values. Trained on control-experiment data that mix multiple experimental settings, the proposed model stabilized an unstable system through feedback control. This is difficult to achieve by fine-tuning existing time-series foundation models, and control became possible only with the introduction of BDT. Ongoing work applying these results to a wide range of target systems is also beginning to bear fruit.
\end{abstract}

%% file: sections/introduction.tex
\section{Introduction}\label{sec:introduction}

In real-world control, data on the input and output signals of a plant accumulate through identification experiments, controller tuning, and routine operation.
Efforts have been made to exploit such data for new identification and controller tuning~\citep{Bittencourt2015NormalOperatingData,Campi2002VRFT}.
Reinforcement-learning-based control has likewise evolved toward offline reinforcement learning, which uses experience gained in past trials to learn policies~\citep{Levine2020OfflineRL}.
Data obtained from diverse experiments and operations should be treated as a common training resource and exploited for prediction, estimation, and control beyond the purposes for which they were originally acquired.

Reusing such data requires capturing the connections under which each time-series signal was generated and how it is used.
In system identification, which estimates system parameters from the input and output signals of a system, it has long been emphasized that the difference in signal generation between open loop and closed loop affects the conditions for estimating the plant dynamics~\citep{Forssell1999ClosedLoop}.
This discussion has been extended to systems of multiple interconnected components, and further to settings in which only some of the internal signals can be measured~\citep{VanDenHof2013DynamicNetworks,Dankers2016PredictorInputSelection}.
Formulations that use experimental data directly for identification, prediction, and control have also been developed for the experimental conditions of individual linear time-invariant systems~\citep{Markovsky2017MissingData,vanWaarde2020DataInformativity}.
Formulations have thus been developed for each specific experimental condition and purpose.
However, a unified formulation that handles experimental data collected under different conditions and the design problems of prediction and control within a single framework has yet to be established.

To address this issue, we extend the data space from time series of signal values to dynamic graphs based on block diagrams, and represent the design problems of prediction and control in a unified manner.
This space consists of dynamic graphs in which each signal component is a vertex, the connections between signals are directed edges defined at each time step, and each vertex carries its signal values and whether they are missing.
On this space, we define an operation that specifies the prediction targets and the known signals, and treat prediction and control as a common masked prediction problem.
This allows us to construct a Control Foundation Model (CFM) that jointly learns response prediction and control-input generation from experimental data with different connections and recorded signals. CFM aims to perform prediction and control without additional training, according to the connections and purpose of each experiment.

To build a predictor based on this formulation, we leverage time-series foundation models pretrained on diverse series.
Time-series foundation models have been developed as a framework for predicting new series from given histories and covariates without additional training~\citep{Ansari2025Chronos2}.
For the inter-signal operations of this predictor, we construct the Block Diagram Transformer (BDT), which reflects the block diagram at each time step.
BDT processes the history of each signal with attention along the time axis, and along the signal axis restricts the attention scope according to the block-diagram connections. By keeping the same parameter configuration as the corresponding time-series foundation model, BDT inherits its pretrained parameters and uses them for predictions that reflect the connection structure.

In experiments on an unstable linear system, we confirmed that our approach improved the prediction accuracy of control inputs compared with fine-tuning a time-series foundation model.
In feedback control, we also obtained a decaying response close to that of linear-quadratic-Gaussian (LQG) control. This work unifies a data representation that preserves differences in experimental conditions with an operation that specifies the design problems of prediction and control, providing a framework for jointly learning prediction and control from diverse experimental data.

The contributions of this paper are threefold:

\begin{itemize}
\item Unified formulation of experimental data and prediction/control problems. We define a data space of dynamic graphs containing connections, signal values, and missing values, together with an operation on this space that specifies the intended use, thereby generalizing masked prediction on time series. This allows response prediction and control-input generation using data from different control experiments to be treated as a common learning problem.
\item Architecture of the control foundation model. We construct BDT, which combines attention along the time axis with graph operations based on block diagrams. BDT inherits the pretrained parameters of the corresponding time-series foundation model while reflecting the connections at each experimental stage in its predictions.
\item Demonstration on an unstable linear system. Experiments on a single unstable linear system confirmed improved prediction accuracy of control inputs and a decaying response close to that of LQG control in feedback control.

\end{itemize}

%% file: sections/related-works.tex
\section{Related Works}\label{sec:related-works}

\subsection{Generalization of time-series models}

Time-series foundation models have been developed as a framework in which a predictor is pretrained on diverse series and applied to unseen series without additional training~\citep{Das2024TimesFM,Ansari2024Chronos,Woo2024Moirai}.
This is achieved by keeping the pretrained parameters fixed while capturing series-specific patterns of variation from the observed history through attention along the time axis and reflecting them in the predictions.
The recently proposed Chronos-2 and TimesFM-3 perform multivariate forecasting that exploits information across series by combining attention along the time axis with attention along the channel axis~\citep{Ansari2025Chronos2,Jain2026TimesFM3}.
Time-series foundation models are thus evolving to exploit not only the temporal variation of individual series but also the relationships among series.
In the control field, studies that perform response prediction and state estimation as extensions of system identification have also been proposed~\citep{Forgione2023ICLSysId,Busetto2024StateEstimators}.

Meanwhile, query-based designs, which select the time steps and variables to be predicted at inference time and handle different inference tasks with the same model, have also been studied.
These models, notably UniMASK and MTM, use a masked-learning framework to select the time steps and channels to be predicted and flexibly predict them~\citep{Carroll2022UniMASK,Wu2023MTM}.
This modeling has led to time-series prediction models that handle covariates and prediction targets in a unified manner.
As an application of such models, M${}^3$PC performs model predictive control that combines the generation of action candidates with the prediction of their outcomes to evaluate and select candidates~\citep{Wen2025M3PC}.
However, these approaches model only a single system, and their generalization to unseen systems has not been demonstrated.

\subsection{Time-series prediction using graph structures}

For traffic and sensor networks, the problem of predicting future values from the connection structure and the observation history at each site has been studied, where the network is represented as a graph with observation sites as nodes and their connections as edges.
For this problem, models that combine temporal feature extraction with inter-site information aggregation by graph neural networks have been proposed~\citep{Yu2018STGCN,Guo2019ASTGCN}.
More recently, research on foundation models, including OpenCity, has pretrained spatiotemporal representations that account for adjacency structures shared across multiple systems, aiming at prediction transfer and adaptation to different systems~\citep{Li2024OpenCity,Zhong2025FactoST}.
Furthermore, query-based designs that specify the time steps and sites to be inferred and handle tasks such as prediction and imputation with a common model have also been studied~\citep{Yuan2025UrbanDiT}.
However, while these models are highly flexible along the graph dimension, they have a fixed prediction length along the time dimension, and their generalization along the time dimension remains unclear.

In this work, we extend the framework of learning from time-series data of different systems and performing prediction and control with the same model to control experiments whose connection structure changes over time.
We represent experimental data with different sequence lengths and numbers of signals as dynamic graphs based on block diagrams, and handle response prediction and control-input generation in a unified manner, including differences in connections and recorded signals.

%% file: sections/background.tex
\section{Background}\label{sec:background}
\begin{figure}[t]
\centering
\includegraphics[width=\linewidth,height=.66\textheight,keepaspectratio]{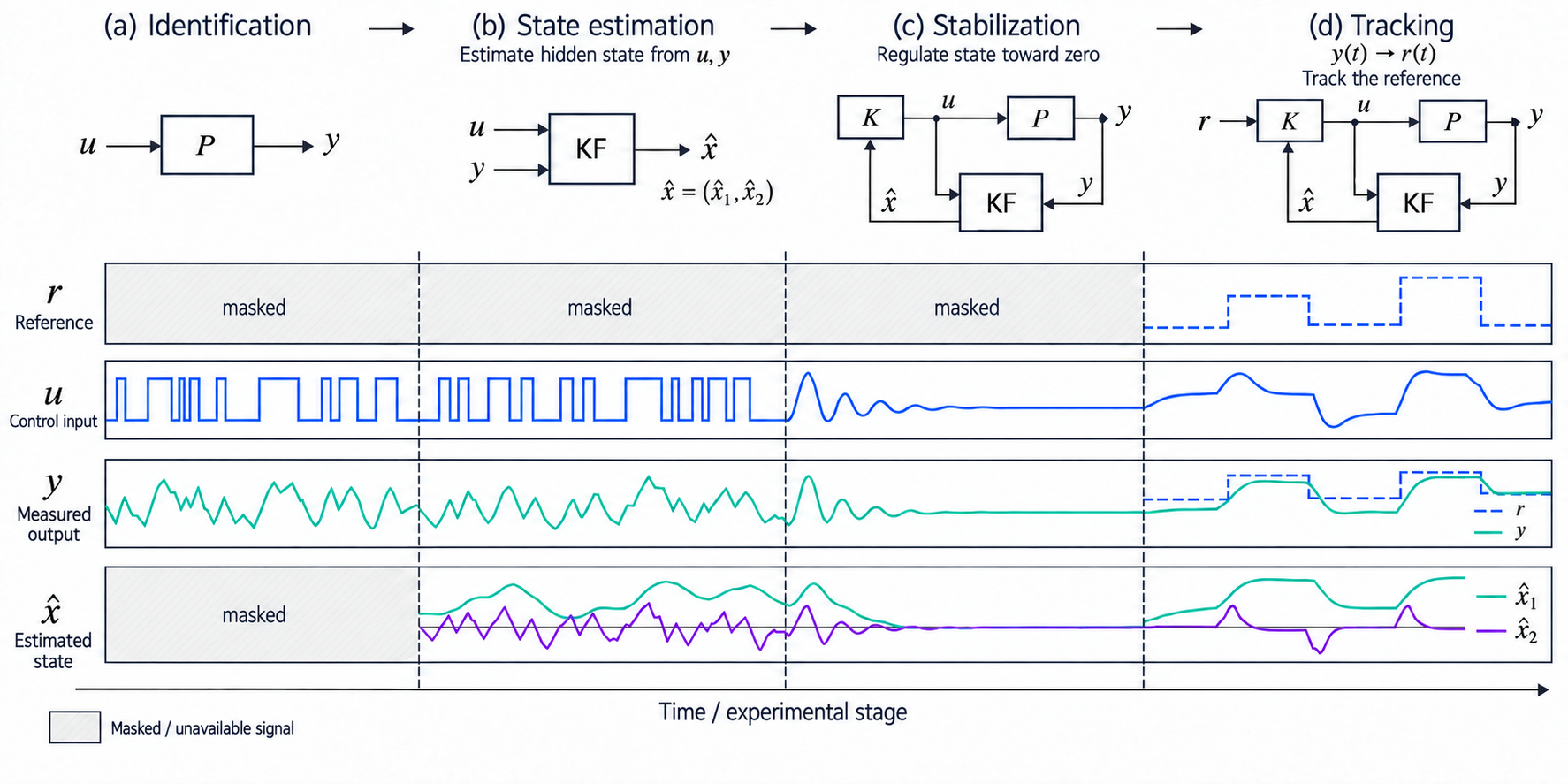}
\caption{Changes in signal connections and available data across experimental stages.
The top row shows the block diagrams for identification, state estimation, stabilization, and tracking, and the bottom row shows schematic time series corresponding to each stage.
\(P\), \(K\), and KF denote the plant, the controller, and the Kalman filter, respectively.
Vertical dashed lines indicate stage boundaries, and shading indicates signal components that cannot be observed.}
\label{fig:cascade}
\end{figure}

Let \(a_s\in\mathbb{R}^{n_a}\) denote the prediction targets and \(b_s\in\mathbb{R}^{n_b}\) the covariates.
We define a model that predicts the \(H\) points starting at time \(t\) from the history of the preceding \(L\) points and the known future covariates as follows:

\begin{equation}\label{eq:bg1}
p_\theta\!\left(a_{t:t+H-1}\mid a_{t-L:t-1},\,b_{t-L:t-1},\,b_{t:t+H-1}\right)
\end{equation}

By pretraining the shared parameters \(\theta\) on diverse series, time-series foundation models can predict future signal values from past history data in a zero-shot manner.

Recent time-series foundation models~\citep{Ansari2025Chronos2} embed the time-series data \(a_{*}, b_{*}\) and their masks patch by patch, and predict the future by combining Transformers along both the time axis and the channel axis.
Let \(h^l_{i,t}\) denote the hidden state of channel \(i\) at patch position \(t\). The attention part of a Transformer layer has the following structure:

\begin{equation}\label{eq:bg2}
\begin{aligned}
\left(h^{l+1/2}_{i,1},\ldots,h^{l+1/2}_{i,L+H}\right)
&=\operatorname{TimeAttn}\!\left(h^l_{i,1},\ldots,h^l_{i,L+H}\right),
&& i\in\{1,\ldots,n_a + n_b\},\\
\left(h^{l+1}_{1,t},\ldots,h^{l+1}_{n_a+n_b,t}\right)
&=\operatorname{GroupAttn}\!\left(h^{l+1/2}_{1,t},\ldots,h^{l+1/2}_{n_a+n_b,t}\right),
&& t\in\{1,\ldots,L+H\}.
\end{aligned}
\end{equation}

This mechanism makes the prediction model applicable to variable dimensions of the prediction targets and covariates.
In what follows, we write \(T\triangleq L+H\) for the combined length of the input (history) range and the prediction range, and \(n\triangleq n_a + n_b\) for the combined dimension of the prediction targets and covariates.

The key question here is what the prediction targets are and what the covariates are in control engineering.
Prediction and control-input generation in control can be viewed uniformly as conditional prediction, which infers unknown signals from known signals.
Treating the input as known and inferring the observations yields response prediction, whereas treating the observation history and the reference values as known and inferring the input yields control-input generation.
In other words, instead of describing the two with separate input-output formats, both can be expressed by switching between known components and prediction targets on a common time series~\citep{Wu2023MTM}.
To express this unification, we collect the time series of all signals into \(x\in\mathbb{R}^{T\times n}\), and let \(M\) denote the index set representing the query (specification of prediction targets) and \(M^C\) its complement, which specifies the known components.
Prediction and control are described as different settings of \(M\) under the same conditional prediction model and training objective:

\begin{equation}\label{eq:bg3}
\begin{aligned}
\text{Prediction model:}\quad
&p_\theta(x_M\mid x_{M^C}),\\
\text{Loss function:}\quad
&\mathrm{Loss}(\theta)
\triangleq
\mathbb{E}_{(x,M)\sim\mathcal{D}}
\left[-\log p_\theta(x_M\mid x_{M^C})\right].
\end{aligned}
\end{equation}

Here, \(M\subseteq\{1,\ldots,T\}\times\{1,\ldots,n\}\), and \(x_M\) and \(x_{M^C}\) denote the prediction targets and the known components, respectively.
\(\mathcal{D}\) is the training distribution over pairs of a time series \(x\) and a prediction-target specification \(M\).

In control-system design, however, the results of earlier stages are carried over to later ones: the plant dynamics are identified, unobservable states are estimated, and this information is then used for stabilization and tracking (Fig.~\ref{fig:cascade}).
In this process, not only the prediction targets but also the mechanisms that process the signals and their connections change.
For example, the input \(u\) during identification is applied externally without using the observation \(y\), whereas it is determined using the estimated state \(\hat{x}\) obtained from the observations in stabilization, and additionally the reference signal \(r\) in tracking.
The connections between these signals and the mechanisms that process them are represented by block diagrams (top row of Fig.~\ref{fig:cascade}).
In this work, we treat the connections at each experimental stage as block diagrams together with the time-series data, and formulate a foundation model for prediction and control across different stages.

%% file: sections/method.tex
\section{Method}\label{sec:method}

\begin{figure}[t]
\centering
\includegraphics[width=\linewidth,height=.66\textheight,keepaspectratio]{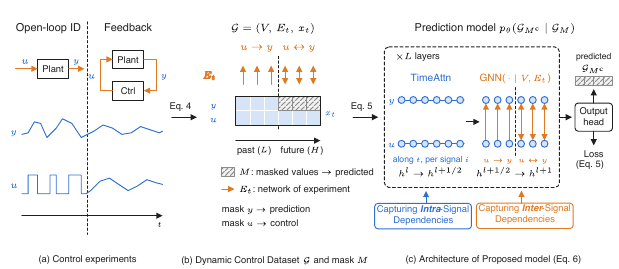}
\caption{Unified representation of control time-series data using block diagrams, and the control foundation model.}
\label{fig:overview}
\end{figure}

\subsection{Unifying control time-series data with block diagrams}

We formulate the block diagram described in the previous section as a graph that represents the connections between signals.
Each signal component is a vertex, and the connection through which one signal is fed into a mechanism that generates another signal is represented as a directed edge.
We associate the signal values at each time step with the vertices and describe changes in the connection structure across experimental stages as changes in the edge set.
We thereby define control time-series data, which treat block diagrams and signal values jointly, as the following dynamic graph.

\begin{definition}[Control time-series data with block diagrams]

We define control time-series data, comprising the connection structure between signals, the signal values, and whether each value is missing, as the following dynamic graph:

\begin{equation}\label{eq:method1}
\mathcal{G}=(V,E_t,X_t)_{t=1}^{T},
\qquad
E_t\subseteq V\times V,
\qquad
X_t=(x_t,\eta_t)\in\mathbb{R}^{|V|}\times\{0,1\}^{|V|}.
\end{equation}

Here, \(V\) is the set of vertices corresponding to the signal components, and \(E_t\) is the set that represents each connection from signal \(i\) to signal \(j\) as a directed edge \((i,j)\).
\(x_t\) denotes the signal values and \(\eta_t\) the missingness mask; \(\eta_{t,i}=1\) indicates that the value is recorded, and \(\eta_{t,i}=0\) indicates that it is missing.

\end{definition}
This formulation allows experimental stages that differ in connection structure and recorded signals to be represented over a common vertex set \(V\).
Open-loop identification has the connection \(u\to y\), whereas tracking control based on state estimation additionally has \(u,y\to\hat{x}\), corresponding to the inputs to the state estimator, and \(r,\hat{x}\to u\), corresponding to the inputs to the controller (top row of Fig.~\ref{fig:cascade}).
Here, a connection between vector signals denotes the collection of edges between their corresponding components.
The estimated state \(\hat{x}\) and the reference signal \(r\), which are not recorded during the identification stage, retain their vertices and are represented by setting the missingness mask to \(\eta_{t,i}=0\) (bottom row of Fig.~\ref{fig:cascade}).
In this way, differences in connection structure and missing signals are represented by \(E_t\) and \(\eta_t\), respectively, so that the data from all stages are handled in a common format.
Below, we define the prediction model and the training objective for the control time-series data \(\mathcal{G}\).

\subsection{Prediction model and loss function with block diagrams}

With this unified representation of control time-series data, prediction problems are unified in a form that accounts for experimental settings that time-series foundation models could not represent.
As with time-series data, we denote the query that specifies the prediction targets by \(M\subset \{1,\ldots, T\}\times V\) and define the unified prediction model and its loss function as follows:

\begin{equation}\label{eq:method2}
\begin{aligned}
\text{Prediction model:}\quad &p_\theta(\mathcal{G}_M\mid\mathcal{G}_{M^C}),\\
\text{Loss function:}\quad &\mathrm{Loss}(\theta)\triangleq\mathbb{E}_{(\mathcal{G},M)\sim\mathcal{D}}\left[-\log p_\theta(\mathcal{G}_M\mid\mathcal{G}_{M^C})\right].
\end{aligned}
\end{equation}

Here, \(\mathcal{G}_M\) and \(\mathcal{G}_{M^C}\) denote data that retain the vertex set and the connection information at each time step and contain the signal values corresponding to \(M\) and \(M^C\), respectively.
The prediction targets are the signal values corresponding to \(M\); the known signal values in \(M^C\) and the block diagram are given as the conditions for prediction.

By defining masking operations and predictive distributions on control time-series data that include the connection structure, this formulation extends the unified treatment of prediction and control to experimental stages with different connection structures.
We refer to a model that pretrains shared parameters on diverse control time-series data and addresses prediction and control problems within this framework as a \textbf{control foundation model}.
Conventional time-series prediction is included as a special case of the control foundation model.

\begin{theorem}

If the connection structure and the time-series variables are independent, Eq.~\eqref{eq:method2} reduces to Eq.~\eqref{eq:bg3}.

\end{theorem}
\begin{proof}

In this proof, we omit the notation for the missingness indicator, which is common to both models.
By the independence of the time-series variables and the connection structure,

\[
\begin{aligned}
p_\theta(\mathcal{G}_M\mid\mathcal{G}_{M^C})
&=\frac{p_\theta(x,V,E_{1:T})}{p_\theta(x_{M^C},V,E_{1:T})}\\
&=\frac{p_\theta(x)\,p_\theta(V,E_{1:T})}{p_\theta(x_{M^C})\,p_\theta(V,E_{1:T})}\\
&=p_\theta(x_M\mid x_{M^C}).
\end{aligned}
\]

Therefore, Eq.~\eqref{eq:method2} reduces to Eq.~\eqref{eq:bg3}.
\end{proof}

\subsection{Architecture of the Block Diagram Transformer}

We propose the \textbf{Block Diagram Transformer (BDT)} as a concrete architecture that realizes the control foundation model.
BDT processes the history of each signal with attention along the time axis and, along the channel axis, uses a graph neural network (GNN) based on the block diagram at each time step.
By replacing the Group Attention in Eq.~\eqref{eq:bg2} with this graph operation, BDT reflects the connection structure in the information exchange between signals.

\begin{equation}\label{eq:method3}
\begin{aligned}
\left(h^{l+1/2}_{i,1},\ldots,h^{l+1/2}_{i,T}\right)
&=\operatorname{TimeAttn}\!\left(h^l_{i,1},\ldots,h^l_{i,T}\right),
&& i\in\{1,\ldots,|V|\},\\
\left(h^{l+1}_{1,t},\ldots,h^{l+1}_{|V|,t}\right)
&=\operatorname{GNN}\!\left(h^{l+1/2}_{1,t},\ldots,h^{l+1/2}_{|V|,t}\mid (V,E_t)\right),
&& t\in\{1,\ldots,T\}.
\end{aligned}
\end{equation}

This architecture includes the attention structure of conventional time-series foundation models as a special case.

\begin{theorem}

Let the GNN in Eq.~\eqref{eq:method3} be a Graph Transformer that uses the same parameters as the Group Attention in Eq.~\eqref{eq:bg2}, with only an attention-scope restriction by a graph mask added.
Then, Eq.~\eqref{eq:bg2} is a special case of Eq.~\eqref{eq:method3} in which the graph \((V,E_t)\) at every time step is the complete graph.

\end{theorem}
\begin{proof}

We define the graph mask corresponding to the connection structure at time \(t\) as

\[
(B_t)_{ij}=
\begin{cases}
0,&(j,i)\in E_t\ \text{or}\ i=j,\\
-\infty,&\text{otherwise}.
\end{cases}
\]

Letting \(H\) be the matrix whose rows are the hidden states \(h_i^\top\), the operation of each head is given by

\[
\operatorname{GraphAttn}(H;B_t)
=
\operatorname{softmax}_{\mathrm{row}}
\left(
\frac{(HW_Q)(HW_K)^\top}{\sqrt{d_k}}+B_t
\right)HW_V,
\]

where \(W_Q,W_K,W_V\) are weight matrices and \(d_k\) is the key dimension.
For the complete graph, \(B_t=0\) at all time steps, so the operation of each head coincides with that of the corresponding head of Group Attention.
Because the operations other than the attention-scope restriction are also shared, the entire operation along the channel axis coincides with Group Attention.
Since the attention along the time axis is shared as well, Eq.~\eqref{eq:method3} coincides with Eq.~\eqref{eq:bg2}.
\end{proof}

In this architecture, Graph Attention is an operation that modifies only the attention mask of Group Attention, and the parameter structure is shared.
Therefore, the pretrained parameters of the corresponding time-series foundation model can be transferred directly to BDT.

In the following, we use this improved model to conduct numerical experiments on prediction problems specific to control engineering.

%% file: sections/results.tex
\section{Results}\label{sec:results}

Even for the same input and output signals, the input is applied externally during identification but is generated from observations during control.
Using trajectories that span identification through control of a single unstable linear system, we evaluated response prediction, input prediction, and feedback control with the predicted inputs.
We compared pretrained Chronos-2, its fine-tuned version, and CFM implemented with BDT.
CFM was given the known block diagrams corresponding to the identification and control segments, together with the switching time.

Fine-tuned Chronos-2 and CFM started from the same pretrained weights.
To preserve the relative scale of inputs and outputs in the model inputs, both were trained without per-channel input/output normalization.
Below, we report the CFM configuration that achieved the smallest root mean square (RMS) of the closed-loop output on the test trajectories, and fine-tuned Chronos-2 with the matched learning rate and maximum number of updates.
These reported configurations were selected after inspecting the test results.
The experimental and training procedures and the results for all search configurations are summarized in \hyperref[sec:appendix]{Appendix~\ref{sec:appendix}}.

\subsection{Accuracy of input and output prediction}

We predicted future outputs or inputs from the histories of the stored test trajectories.
For response prediction, the known future inputs were also given; for input prediction, only past inputs and outputs were given.
CFM handled both tasks by changing the specification of known components and prediction targets.

In output prediction, fine-tuned Chronos-2 and CFM achieved nearly the same accuracy, whereas in input prediction, CFM achieved a smaller error (Table~\ref{tab:prediction}).
The two models, which are close in response-prediction accuracy, differed when inferring the inputs.

\begin{table}[htbp]
\centering
\caption{Multi-step prediction performance. Mean \(\pm\) sample standard deviation of the per-trajectory root-mean-square error (RMSE) over 100 test trajectories.
The input-prediction statistics include both the identification segment with the M-sequence and the LQG control segment.}
\label{tab:prediction}
\begingroup\small
\begin{tabularx}{\linewidth}{@{}p{.27\linewidth}XX@{}}
\toprule
Method & Output-prediction RMSE & Input-prediction RMSE\\
\midrule
Chronos-2 & 0.4235 \(\pm\) 0.2507 & 1.8814 \(\pm\) 0.8008\\
fine-tuned Chronos-2 & \textbf{0.0289 \(\pm\) 0.0042} & 1.6793 \(\pm\) 0.7333\\
CFM & 0.0290 \(\pm\) 0.0035 & \textbf{0.7772 \(\pm\) 0.0271}\\
\bottomrule
\end{tabularx}
\endgroup
\end{table}

\subsection{Feedback control with predicted inputs}

Next, we performed feedback control that drives the system to the origin.
At each time step, the model predicted an input sequence, and its first element was applied to the plant.
Because the resulting observation is appended to the history before the next prediction, each model uses, after control starts, a history generated by its own inputs.

\begin{figure}[t]
\centering
\includegraphics[width=\linewidth,height=.66\textheight,keepaspectratio]{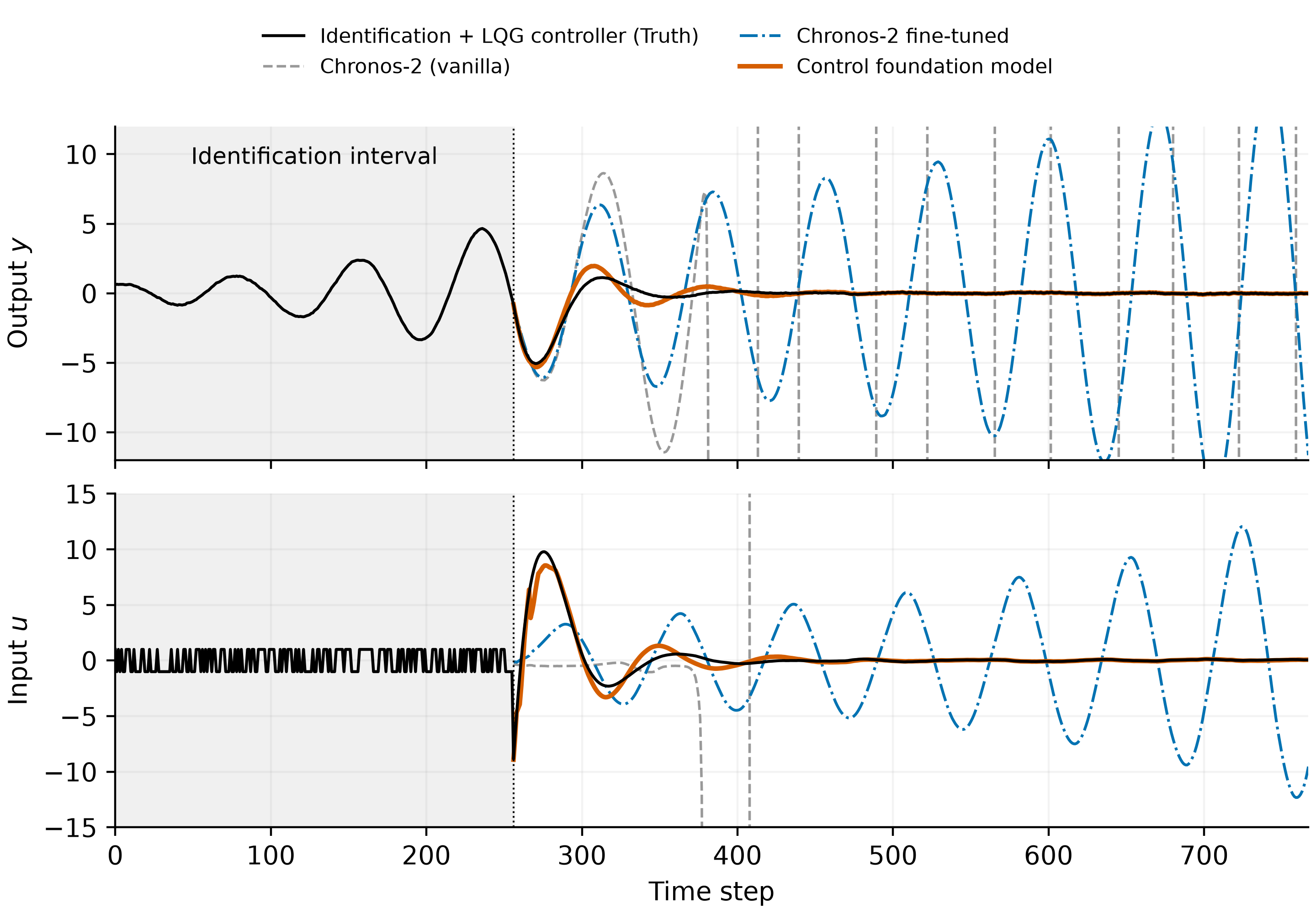}
\caption{Feedback control response with predicted inputs. Output $y$ (top) and input $u$ (bottom) of a representative trajectory.
Shading indicates the identification segment.
LQG is a reference that uses the known model and the observation at the current time step, whereas the learned models use observations up to the previous time step.
The displayed range is limited to compare the decaying responses, but the evaluation metrics use all time points, including those outside this range.
Full-range plots are shown in \hyperref[sec:full-range-response]{Appendix~\ref{sec:appendix}}.}
\label{fig:control}
\end{figure}

On a representative trajectory starting from the same identification history and the same state at the start of control, CFM exhibited a decaying response close to the LQG reference, whereas fine-tuned Chronos-2 retained large-amplitude oscillations (Fig.~\ref{fig:control}).
Across all six configurations with matched learning rates and maximum numbers of updates, CFM also achieved a smaller closed-loop output RMS aggregated over all test trajectories (\hyperref[tab:matched-control-results]{Table~\ref{tab:matched-control-results} in Appendix~\ref{sec:appendix}}).

These results indicate that reflecting the stage-dependent connections in prediction is effective for extending time-series foundation models from response prediction to control-input generation.

%% file: sections/conclusion.tex
\section{Conclusion}\label{sec:conclusion}

We formulated prediction and control as a common prediction problem by introducing dynamic graphs that jointly represent the connections between signals, signal values, and missingness, together with an operation that specifies known signals and prediction targets. This formulation provides a foundation for treating control experiments with different connections and recorded signals as a common training resource. Its concrete architecture, the Block Diagram Transformer (BDT), inherits the pretrained parameters of a time-series foundation model while incorporating block-diagram connections into cross-series attention.

In experiments on a single unstable linear system, the proposed model reduced the control-input prediction error and the closed-loop output RMS while maintaining response-prediction accuracy comparable to that of a fine-tuned time-series foundation model. The significance of this result is most evident in closed-loop control simulations: whereas the closed-loop system with the existing model failed to stabilize the unstable plant, the closed-loop system with the proposed method achieved stabilization. These results indicate that handling stage-dependent connection structures is effective for extending time-series prediction to control-input generation.

This work has already been extended to unseen system instances, multi-input multi-output (MIMO) systems, and joint learning from linear systems and rigid-body arms, and we are evaluating its generalization across different systems and signal dimensions. These studies examine to what extent learning from multiple experiments can compensate for information lacking in individual experiments, and the range over which response prediction, input prediction, and closed-loop control can be performed without additional training. Our framework aims to reuse control-experiment records beyond the purposes for which they were acquired and to learn prediction and control jointly.

%% file: sections/appendix.tex
\section{Details of experiments and evaluation on a single unstable system}\label{sec:appendix}

\subsection{Experimental system, training, and evaluation procedure}\label{sec:data}

We generated trajectories from a single unstable linear system and split them into training, validation, and test sets at the trajectory level.

The state and observation equations are

\begin{equation}\label{eq:app1}
x_{t+1}=A_dx_t+B_du_t+w_t,\qquad y_t=Cx_t+Du_t+v_t
\end{equation}

with state dimension 3, one input and one output, and $D=0$. The continuous-time system was discretized by zero-order hold with step size $\Delta t=0.1$. The fixed system ID is 967; the largest real part of the eigenvalues of the continuous-time matrix is approximately $+0.0744768$, and the spectral radius of the discrete-time matrix is approximately 1.0074755. The plant is open-loop unstable.

All trajectories use the same $A_d,B_d,C,D$, while the initial state, excitation phase, and noise vary.

\begin{center}
\begingroup\normalsize
\setlength{\tabcolsep}{3pt}
\begin{tabularx}{\linewidth}{@{}p{.27\linewidth}X@{}}
\toprule
Item & Setting\\
\midrule
Number of trajectories and split & 1000 trajectories: train 800, validation 100, test 100. The stored ID assignment is used\\
Initial state & $x_0\sim\mathcal N(0,I_3)$\\
Process noise & $w_t\sim\mathcal N(0,10^{-4}I_3)$\\
Observation noise & $v_t\sim\mathcal N(0,10^{-4})$\\
Train/validation length & 512 points, $t=0,\ldots,511$\\
Stored test length & 768 points, $t=0,\ldots,767$\\
Identification input & 9-stage M-sequence with amplitude $\pm1$, period 511, SciPy \texttt{max\_len\_seq(9,taps=[5])}, phase varied per trajectory\\
Switching time for train/validation & Selected per trajectory from $\{16,32,\ldots,496\}$\\
Switching time for test & 256 for all trajectories\\
System seed & 2026091801, fixed ID 967\\
Trajectory seed & SeedSequence generated from 2026091802 and the trajectory ID\\
Switch seed & 2026091702\\
M-sequence phase seed & 2026091703 and the trajectory ID\\
Test extension noise seed & 2026091704 and the trajectory ID\\
\bottomrule
\end{tabularx}
\endgroup
\end{center}

The teacher trajectories for training are excited by the M-sequence before the switch and controlled by known-model LQG after the switch. The LQR weights are $Q=I_3,R=1$, and the steady-state Kalman filter uses the noise covariances above. The teacher LQG determines $u_t$ from the estimated state that incorporates the current observation $y_t$. In contrast, closed-loop control by the learned models determines $u_t$ from observations up to $t-1$. Known-model LQG and the learned models therefore differ in both model knowledge and observation timing.

The storage order is $(u,y)$. Simulation arrays are in double precision, and the training signals stored in Arrow are float32. The stored NPZ files hold the matrix values and noise sequences in double precision, and the manifest holds the split information.

The matrices of the discrete-time plant used are shown below (displayed values are rounded).

\begin{equation}\label{eq:app2}
A_d=\begin{bmatrix}
1.0142272252 & 0.0615079350 & -0.0559719638\\
-0.0196168887 & 0.8987286319 & -0.0643002095\\
0.0685985842 & 0.0450867269 & 1.0081244900
\end{bmatrix},\qquad
B_d=\begin{bmatrix}-0.0035569573\\0.0852462557\\-0.0229012672\end{bmatrix},
\end{equation}

\begin{equation}\label{eq:app3}
C=\begin{bmatrix}0.2732344408 & 0.0906551202 & -0.5403868591\end{bmatrix},\qquad D=0.
\end{equation}

\subsubsection{Prediction tasks and connection structure}\label{sec:training}

For each origin $c$, the observed history $u_{0:c-1},y_{0:c-1}$ is given as input, and 16 points are predicted. We use the expanding history up to the origin rather than a fixed-length sliding window. Short histories within a batch are left-padded with NaN, and only the valid part is passed to the model.

\begin{center}
\begingroup\normalsize
\setlength{\tabcolsep}{3pt}
\begin{tabularx}{\linewidth}{@{}p{.27\linewidth}XX@{}}
\toprule
Task & Known information & Prediction target\\
\midrule
y prediction & Past u/y and future input $u_{c:c+15}$ & $y_{c:c+15}$\\
u prediction & Past u/y only & $u_{c:c+15}$\\
\bottomrule
\end{tabularx}
\endgroup
\end{center}

y prediction, which uses future u, and u prediction, which uses no future covariates, are trained in the same batch. Series other than the prediction targets are masked out of the loss. Because u prediction also includes the M-sequence input in the identification segment, the u errors in the table should not be interpreted as input errors in the control segment alone.

Fine-tuned Chronos-2 and CFM use the same Chronos-2 initial weights and parameter configuration, and all parameters are fine-tuned. Fine-tuned Chronos-2 allows attention between series within a group. CFM applies an inter-series attention mask corresponding to the known block diagram, allowing the path u\(\to\)y in the identification segment and u\(\leftrightarrow\)y in the control segment. Self-paths are retained. The switching time is given in the data; this is not an experiment in which the model estimates the switching time.

In the implementation, y\(\to\)u is allowed for a history patch once its end exceeds the switching time, and for a prediction patch when the prediction origin is at or after the switching time. Inter-series connections at the REG token positions are allowed. Hence, the mask is not simply a uniform static adjacency matrix but depends on the patch position and the known switching time.

\subsubsection{Common training settings}

Each run started from the original Chronos-2 weights. The 6000-step and 12000-step runs are also independent runs; the latter is not a continuation from the 6000-step checkpoint.

\begin{center}
\begingroup\normalsize
\setlength{\tabcolsep}{3pt}
\begin{tabularx}{\linewidth}{@{}p{.27\linewidth}X@{}}
\toprule
Item & Common setting\\
\midrule
Backbone & Chronos-2, 12 layers, 12 heads, hidden size 768, FFN 3072\\
Input patch / stride / output patch & 16 / 16 / 16\\
Prediction length & 16, with REG token\\
Quantiles & 21 levels: 0.01, 0.05, 0.10, 0.15, \(\ldots\), 0.95, 0.99\\
Point prediction & Quantile 0.5\\
Train origins & $64,80,\ldots,496$: 28 origins, 800\(\times\)28=22,400 windows\\
Validation origins & 128,256,384,496: 100\(\times\)4=400 windows\\
Optimizer & AdamW, weight decay 0, default betas=(0.9,0.999), eps=$10^{-8}$\\
LR schedule & cosine, warmup 25 updates, minimum LR=$2\times10^{-6}$\\
Gradient clipping & norm 1.0\\
Dropout / attention dropout & 0 / 0.1; eval mode during evaluation\\
Batch / gradient accumulation & Current batch 16, accumulation 1\\
Validation frequency & Every 100 updates\\
Checkpoint selection & Minimum \texttt{val/target\_loss} within the run\\
Precision and resources & float32, a single GH200 GPU per run\\
Training seed for the current grid & 42, shared across all conditions\\
\bottomrule
\end{tabularx}
\endgroup
\end{center}

The training loss is the pinball loss over the 21 quantiles. The median is used for point prediction.

\subsubsection{Evaluation procedure and performance metrics}\label{sec:metrics}

\paragraph{Evaluation conditions}

Let $N$ be the number of evaluation trajectories, $n_x,n_u,n_y$ the state, input, and output dimensions, and $\Delta t$ the sampling interval. Let $T_{\mathrm{id}}$ be the length of the identification segment, $T_{\mathrm{ctrl}}$ the length of the control evaluation segment, and $T_{\mathrm{all}}=T_{\mathrm{id}}+T_{\mathrm{ctrl}}$ the stored trajectory length. Let $H$ be the prediction length, $\delta$ the re-prediction interval, $T_{\mathrm{pred}}$ the trajectory length used for prediction evaluation, and $\mathcal C$ the set of prediction origins.

\paragraph{Prediction evaluation (Table~\ref{tab:prediction} in the main text)}

At each origin of the stored teacher trajectories, the measured history is given and the future output or input is predicted. Histories generated by the model itself under closed-loop control are not used. The evaluation of input prediction includes both the identification segment with the M-sequence and the LQG control segment.

For task $q\in\{y,u\}$ and trajectory $i$, the RMSE and its mean and sample standard deviation across trajectories are defined as follows.

\begin{equation}\label{eq:app4}
r_{i,q}=\sqrt{\frac{1}{|\mathcal C|H}\sum_{c\in\mathcal C}\sum_{h=0}^{H-1}(\hat q^{(i)}_{c+h\mid c}-q^{(i)}_{c+h})^2},
\end{equation}

\begin{equation}\label{eq:app5}
\bar r_q=\frac{1}{N}\sum_{i=1}^{N}r_{i,q},\qquad
s_q=\sqrt{\frac{1}{N-1}\sum_{i=1}^{N}(r_{i,q}-\bar r_q)^2}.
\end{equation}

Table~\ref{tab:prediction} in the main text reports $\bar r_q\pm s_q$.

\paragraph{Closed-loop evaluation (Fig.~\ref{fig:control} in the main text)}

The same plant, identification history, state at the start of control, and noise sequences are used for every method. The true state is used only to initialize the simulator and is not given to the learned models. At each control time step, the future input sequence is predicted from past inputs and outputs, and the first input is applied to the environment. The resulting observation is appended to the history, and re-prediction is performed at the next time step. After control starts, each model uses the history produced by its own control. The objective is regulation to the origin. The design and observation timing of the LQG reference are described in \hyperref[sec:data]{"Experimental system, training, and evaluation procedure" (Section~\ref{sec:data})}.

The magnitudes of the output and input are evaluated by the RMS aggregated over all trajectories, all control time steps, and all components.

\begin{equation}\label{eq:app6}
J_q=\left[\frac{1}{NT_{\mathrm{ctrl}}n_q}\sum_{i=1}^{N}\sum_{t=T_{\mathrm{id}}}^{T_{\mathrm{all}}-1}\lVert q_t^{(i)}\rVert_2^2\right]^{1/2},\qquad q\in\{y,u\}.
\end{equation}

As an auxiliary metric, we define the deviation from the corresponding LQG reference.

\begin{equation}\label{eq:app7}
E_q=\left[\frac{1}{NT_{\mathrm{ctrl}}n_q}\sum_{i=1}^{N}\sum_{t=T_{\mathrm{id}}}^{T_{\mathrm{all}}-1}\lVert q_t^{(i)}-q_{t,\mathrm{LQG}}^{(i)}\rVert_2^2\right]^{1/2}.
\end{equation}

$J_q$ represents the magnitude relative to the origin, and $E_q$ the difference from the LQG reference. In both cases, squared values are aggregated over all trajectories before taking the square root.

\paragraph{Specific values in this experiment}

\begin{center}
\begingroup\normalsize
\setlength{\tabcolsep}{3pt}
\begin{tabularx}{\linewidth}{@{}p{.27\linewidth}X@{}}
\toprule
Variable / setting & Value\\
\midrule
$N$ & 100\\
$(n_x,n_u,n_y)$ & $(3,1,1)$\\
$\Delta t$ & 0.1\\
$T_{\mathrm{pred}}$ & 512\\
$\mathcal C$ & $\{64,80,\ldots,496\}$ (28 origins)\\
$H$ & 16\\
Prediction evaluation range & $t=64,\ldots,511$ of each trajectory (448 points)\\
$T_{\mathrm{id}}$ & 256\\
$T_{\mathrm{ctrl}}$ & 512\\
$T_{\mathrm{all}}$ & 768\\
$\delta$ & 1\\
Trajectory in Fig.~\ref{fig:control} of the main text & Trajectory ID 1\\
\bottomrule
\end{tabularx}
\endgroup
\end{center}

The prediction length $H$ was fixed across all conditions. No saturation was applied to the control input.

\subsection{Reasons for removing per-channel normalization}\label{sec:normalization}

If the input and output are standardized independently for each history, the relative scale between the signals can no longer be recovered from the standardized signal values alone.
Let $\mu_u,\mu_y$ be the means of the respective series over the same observed history segment and $s_u,s_y$ their positive standard deviations. Then

\[
\tilde u=\frac{u-\mu_u}{s_u},\qquad
\tilde y=\frac{y-\mu_y}{s_y}
\]

holds.
As a simple example, the proportional relation $u=ky$ becomes, after standardization,

\[
\tilde u=k\frac{s_y}{s_u}\tilde y+
\frac{k\mu_y-\mu_u}{s_u}.
\]

When non-constant series exactly satisfy $u=ky,\ k>0$ over the same history segment, $\mu_u=k\mu_y$ and $s_u=ks_y$ imply

\[
\tilde u=\tilde y.
\]

In this case, the standardized signal values coincide for different positive gains, so the magnitude of the original proportionality coefficient cannot be distinguished from those values alone.

In this experiment, to preserve the relative scale between the raw signals in the model input, we replaced the input/output processing of fine-tuned Chronos-2 and CFM with the identity map.
The removed processing consists of per-history, per-series standardization, the asinh transform, and the inverse transform of the output; the LayerNorm layers inside the Transformer were retained.
Chronos-2 vanilla used the original input/output processing.

The above is the rationale for the choice of preprocessing; it is not a claim that a full model, including reintroduction of statistics or inverse transformation of the output, cannot handle proportional relations.
In addition, since vanilla and the fine-tuned versions differ simultaneously in whether they were trained and in preprocessing, we do not interpret their performance difference as the effect of removing normalization alone.

\subsection{Results by training configuration}\label{sec:search}

\subsubsection{Search space and selection procedure}

Fine-tuned Chronos-2 and CFM were trained independently over the same search space. Each run started from the original pretrained weights, and the common settings follow \hyperref[sec:training]{"Prediction tasks and connection structure" (Section~\ref{sec:data})}.

\begin{center}
\begingroup\normalsize
\setlength{\tabcolsep}{3pt}
\begin{tabularx}{\linewidth}{@{}p{.27\linewidth}X@{}}
\toprule
Item & Search space\\
\midrule
Learning rate & $\{3\times10^{-5},10^{-4},3\times10^{-4}\}$\\
Maximum number of updates & $\{6000,12000\}$\\
Batch size & 16 (fixed)\\
Number of configurations & 6 per model, 12 in total\\
\bottomrule
\end{tabularx}
\endgroup
\end{center}

Let $\mathcal A$ be the set of CFM configurations, $\mathcal K_a$ the set of saved updates for configuration $a$, and $\theta_{a,k}$ the weights at update $k$. The checkpoint of each run was selected by validation loss, and the adopted configuration was selected by its closed-loop output RMS.

\begin{equation}\label{eq:app8}
k^\star(a)\in\operatorname*{arg\,min}_{k\in\mathcal K_a}\mathcal L_{\mathrm{val}}(\theta_{a,k}),\qquad
 a^\star\in\operatorname*{arg\,min}_{a\in\mathcal A}J_y(\theta_{a,k^\star(a)}).
\end{equation}

$J_y$ was computed on the test trajectories. For fine-tuned Chronos-2 in the main text, we used the same learning rate, maximum number of updates, and batch size as the adopted CFM.

\begin{center}
\begingroup\normalsize
\setlength{\tabcolsep}{3pt}
\begin{tabularx}{\linewidth}{@{}p{.22\linewidth}YYYY@{}}
\toprule
Model & Input/output normalization & Adopted learning rate & Max. updates & Adopted checkpoint\\
\midrule
Chronos-2 vanilla & Yes & --- & --- & Original pretrained weights\\
fine-tuned Chronos-2 & No & $10^{-4}$ & 12000 & Update 10800\\
CFM & No & $10^{-4}$ & 12000 & Update 11800\\
\bottomrule
\end{tabularx}
\endgroup
\end{center}

\subsubsection{Closed-loop comparison under matched training configurations}

\begin{table}[htbp]
\centering
\caption{Closed-loop output RMS under matched training configurations. The checkpoint of each run was selected by validation loss. The maximum number of updates differs from the number of updates of the checkpoint actually selected.}
\label{tab:matched-control-results}
\begingroup\normalsize
\setlength{\tabcolsep}{3pt}
\begin{tabularx}{\linewidth}{@{}p{.22\linewidth}YYY@{}}
\toprule
Learning rate & Max. updates & fine-tuned Chronos-2 & CFM\\
\midrule
3e-05 & 6,000 & 160.37797 & 7.92710\\
3e-05 & 12,000 & 152.49778 & 7.80909\\
0.0001 & 6,000 & 34.22627 & 1.79427\\
0.0001 & 12,000 & 11.04592 & \textbf{1.27467}\\
0.0003 & 6,000 & 12.22290 & 1.34924\\
0.0003 & 12,000 & 10.02856 & 1.74633\\
\bottomrule
\end{tabularx}
\endgroup
\end{table}

\subsubsection{Details of all search configurations}

The evaluation metrics follow \hyperref[sec:metrics]{"Evaluation procedure and performance metrics" (Section~\ref{sec:data})}.

\begin{center}
\begingroup\normalsize
\setlength{\tabcolsep}{3pt}
\begin{tabularx}{\linewidth}{@{}p{.22\linewidth}YYYYYY@{}}
\toprule
Configuration & LR & Max. updates & Min. validation loss & Selected update & Test y RMS & Test u RMS\\
\midrule
fine-tuned Chronos-2 & 3e-05 & 6000 & 8.60341 & 5900 & 160.37797 & 22.85499\\
fine-tuned Chronos-2 & 3e-05 & 12000 & 6.03964 & 9700 & 152.49778 & 24.12714\\
fine-tuned Chronos-2 & 0.0001 & 6000 & 8.50126 & 3700 & 34.22627 & 18.32350\\
fine-tuned Chronos-2 & 0.0001 & 12000 & 6.68355 & 10800 & 11.04592 & 12.09465\\
fine-tuned Chronos-2 & 0.0003 & 6000 & 9.78136 & 3700 & 12.22290 & 5.08303\\
fine-tuned Chronos-2 & 0.0003 & 12000 & 9.09383 & 11900 & 10.02856 & 5.95962\\
CFM & 3e-05 & 6000 & 4.83451 & 6000 & 7.92710 & 30.45850\\
CFM & 3e-05 & 12000 & 2.24564 & 11800 & 7.80909 & 27.96356\\
CFM & 0.0001 & 6000 & 3.63282 & 6000 & 1.79427 & 2.67895\\
CFM & 0.0001 & 12000 & 2.86828 & 11800 & 1.27467 & 2.28126\\
CFM & 0.0003 & 6000 & 3.38390 & 5900 & 1.34924 & 1.97067\\
CFM & 0.0003 & 12000 & 2.72442 & 11700 & 1.74633 & 1.93835\\
\bottomrule
\end{tabularx}
\endgroup
\end{center}

In every configuration matched by learning rate and maximum number of updates, the closed-loop output RMS of CFM was smaller than that of fine-tuned Chronos-2.

The control performance of the adopted CFM and its difference from the LQG reference are shown below.

\begin{center}
\begingroup\normalsize
\setlength{\tabcolsep}{3pt}
\begin{tabularx}{\linewidth}{@{}p{.22\linewidth}YYY@{}}
\toprule
$J_y$ & $J_u$ & $E_y$ & $E_u$\\
\midrule
1.27467 & 2.28126 & 0.79346 & 1.28305\\
\bottomrule
\end{tabularx}
\endgroup
\end{center}

\subsubsection{Input prediction error and closed-loop output RMS for all configurations}

\begin{table}[htbp]
\centering
\caption{Input prediction error and closed-loop output RMS for all 12 configurations. Input prediction is reported as the mean \(\pm\) sample standard deviation of per-trajectory RMSE, evaluated on the same 100 test trajectories and 28 origins as in Table~\ref{tab:prediction}.}
\label{tab:prediction-control-relation}
\begingroup\normalsize
\setlength{\tabcolsep}{3pt}
\begin{tabularx}{\linewidth}{@{}p{.22\linewidth}YYYY@{}}
\toprule
Method & Learning rate & Max. updates & Input prediction RMSE & Closed-loop output RMS\\
\midrule
fine-tuned Chronos-2 & 3e-05 & 6,000 & 1.7012 \(\pm\) 0.7393 & 160.37797\\
fine-tuned Chronos-2 & 3e-05 & 12,000 & 1.5970 \(\pm\) 0.8323 & 152.49778\\
fine-tuned Chronos-2 & 0.0001 & 6,000 & 1.7109 \(\pm\) 0.7382 & 34.22627\\
fine-tuned Chronos-2 & 0.0001 & 12,000 & 1.6793 \(\pm\) 0.7333 & 11.04592\\
fine-tuned Chronos-2 & 0.0003 & 6,000 & 1.7423 \(\pm\) 0.7500 & 12.22290\\
fine-tuned Chronos-2 & 0.0003 & 12,000 & 1.6906 \(\pm\) 0.7297 & 10.02856\\
CFM & 3e-05 & 6,000 & 0.6922 \(\pm\) 0.0158 & 7.92710\\
CFM & 3e-05 & 12,000 & 0.1922 \(\pm\) 0.0500 & 7.80909\\
CFM & 0.0001 & 6,000 & 0.7410 \(\pm\) 0.0226 & 1.79427\\
CFM & 0.0001 & 12,000 & 0.7772 \(\pm\) 0.0271 & 1.27467\\
CFM & 0.0003 & 6,000 & 0.7208 \(\pm\) 0.0210 & 1.34924\\
CFM & 0.0003 & 12,000 & 0.7868 \(\pm\) 0.0274 & 1.74633\\
\bottomrule
\end{tabularx}
\endgroup
\end{table}

\subsubsection{Full-range closed-loop responses}\label{sec:full-range-response}

\begin{figure}[t]
\centering
\includegraphics[width=\linewidth,height=.66\textheight,keepaspectratio]{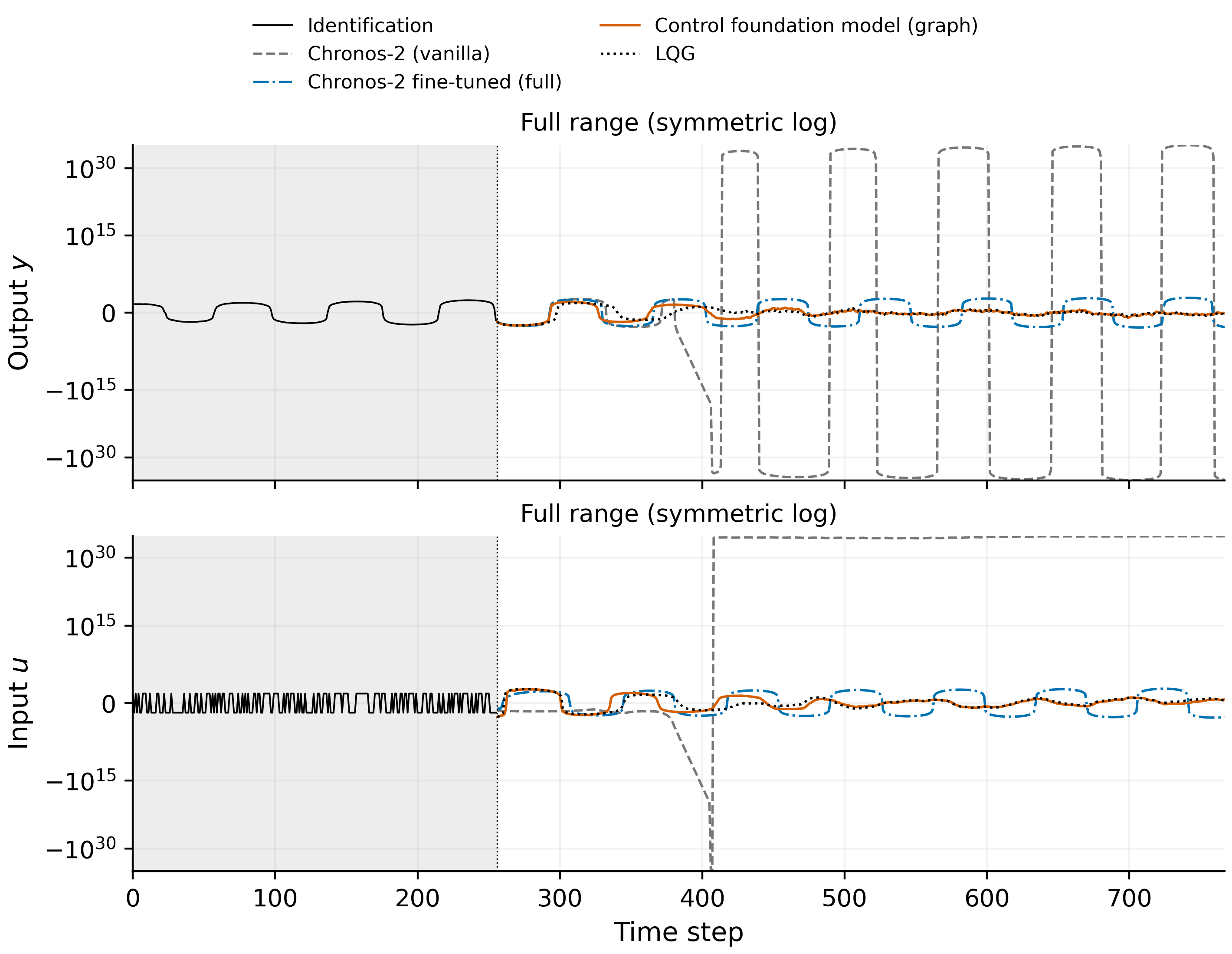}
\caption{Full-range display of the same representative trajectory as in Fig.~\ref{fig:control}. The output (top) and input (bottom) are shown over their full range on a symmetric logarithmic axis in the left column and as a zoomed view on a linear axis in the right column. Gray: Chronos-2 vanilla; blue: fine-tuned Chronos-2; orange: CFM; black: identification and LQG reference response. The display range in the right column does not indicate an input constraint. No input saturation was applied in this experiment.}
\label{fig:control-full-range}
\end{figure}

\subsubsection{Reproducibility information}\label{sec:reproducibility}

The experiment ID is \texttt{single-unstable-mseq-20260918}. The data-generation seeds are given in \hyperref[sec:data]{"Experimental system, training, and evaluation procedure" (Section~\ref{sec:data})}, the training seed and settings in \hyperref[sec:training]{"Prediction tasks and connection structure" (Section~\ref{sec:data})}, and the adopted configurations in \hyperref[sec:search]{"Results by training configuration" (Section~\ref{sec:search})}.

\begin{center}
\begingroup\normalsize
\setlength{\tabcolsep}{3pt}
\begin{tabularx}{\linewidth}{@{}p{.27\linewidth}X@{}}
\toprule
Item & Environment\\
\midrule
GPU for training and current evaluation & NVIDIA GH200, 1 per run\\
PyTorch / CUDA & 2.13 / 13\\
Transformers / Lightning & 5.14.1 / 2.6.6\\
Training precision & float32\\
\bottomrule
\end{tabularx}
\endgroup
\end{center}

Code, data, and settings are summarized in the list of reference materials.

%% file: sections/appendix-c.tex
\section{Evaluating the effect of the graph mask on causal relations}\label{sec:graph-mask-evaluation}

\subsection{Evaluation approach}\label{sec:graph-mask-causal-evaluation}

In predicting control time series, not only the correlations between signals but also the causal direction, i.e., which signals each signal is generated from, is meaningful.
The graph mask restricts the attention scope between series on the basis of this known structure.
In this evaluation, we examine whether the imposed constraints are reflected in the dependencies of the predictions of the model as a whole.

The evaluation fixes all other conditions, modifies only the observed history of some signals, and compares the resulting changes in the predictions.
For signals that, by construction, have no influence on the prediction targets, we check that the predictions are invariant to such modifications.
This test is separate from the magnitude of the prediction error: it checks the consistency between the known causal structure and the dependencies used by the model.
Below, we carry out this evaluation using an identification task as a concrete example.

\subsection{Verification on identification}\label{sec:identification-perturbation-protocol}

As a concrete example, we use $0\le t<2$ s of one SISO identification example from Stage 2.
In this identification task, the generation rule for the input $u$ does not refer to the output $y$, and the causal direction is $u\to y$.
Therefore, for the model in which $y\to u$ is prohibited, we check whether the prediction of $u$ remains unchanged when only the past $y$ is modified.
The verification covers only the identification segment and does not include the control segment.

The signals present in this segment are $u$ and $y$; the reference signal $r$ is absent.
With rows as receivers, columns as sources, and the order $(u,y)$, the additive masks are

\[
B_t^{\mathrm{dynamic}}=
\begin{bmatrix}0&-\infty\\0&0\end{bmatrix},\qquad
B_t^{\mathrm{complete}}=
\begin{bmatrix}0&0\\0&0\end{bmatrix}
\]

and they are constant throughout the segment.
Here, $0$ denotes that attention is allowed and $-\infty$ that it is prohibited; dynamic prohibits $y\to u$.

For the 200 recorded points of the identification segment, we predicted $u_c$ one step ahead at 168 time steps with origins $c=32,\ldots,199$.
The input at each origin is the recorded $u_{0:c-1},y_{0:c-1}$.
Between the two inputs being compared, only the past outputs were changed to

\[
\widetilde y_j=-3y_j+1,\qquad 0\le j<c
\]

while the past $u$, time, graph, missingness and prediction-target specification, and the weights of each model were kept fixed.
Both $u_c$ and $y_c$ at the prediction time were hidden, and predictions were not fed back into the history for the next origin.
This is a comparison of predictions on the same recorded history, not a closed-loop evaluation in which the plant is re-simulated according to the modified $y$.

\subsection{Results and interpretation}\label{sec:identification-perturbation-results}

\begin{figure}[t]
\centering
\includegraphics[width=\linewidth,height=.73\textheight,keepaspectratio]{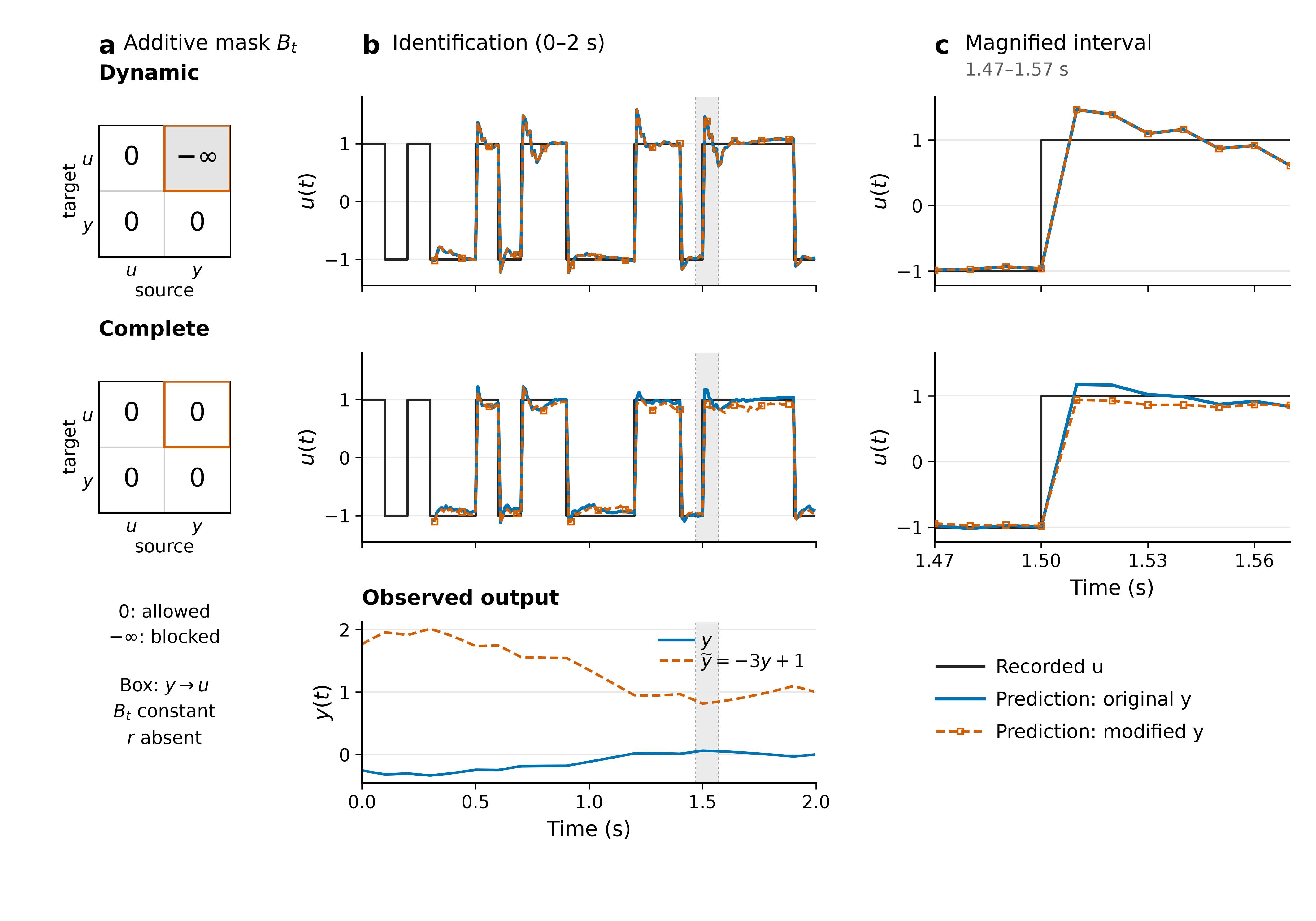}
\caption{Graph mask and the dependence of input prediction on the output history. \textbf{a:} Additive masks for dynamic and complete. The boxes correspond to $y\to u$. \textbf{b:} Recorded $u$ over the entire identification segment (black), prediction with the original $y$ (blue), and prediction with the modified $y$ (orange dashed). The bottom panel shows the original $y$ and $\widetilde y$. \textbf{c:} Zoomed view of $u$ for both models over the same interval, 1.47--1.57 s, with the same vertical axis. The gray band indicates the zoomed range. This range was chosen as 0.05 s on either side of 1.52 s, where the prediction difference of complete is largest within the identification segment, and does not represent a typical interval.}
\label{fig:identification-perturbation}
\end{figure}

\begin{center}
\begingroup\normalsize
\setlength{\tabcolsep}{3pt}
\begin{tabularx}{\linewidth}{@{}p{.27\linewidth}XX@{}}
\toprule
Model & Maximum absolute difference in input prediction before and after modification & Time steps with identical predictions before and after modification\\
\midrule
dynamic & 0 & 168 / 168\\
complete & 0.23722672 & 0 / 168\\
\bottomrule
\end{tabularx}
\endgroup
\end{center}

In this identification example, the predictions of dynamic before and after the modification were byte-identical at all 168 time steps, whereas those of complete changed at every time step.
For dynamic, invariance to modification of the output history was confirmed for the prohibited direction $y\to u$, yielding prediction dependencies consistent with the specified causal direction.
For complete, whose configuration allows dependence in this direction, the prediction of $u$ changed.
The invariance here refers to agreement between the predictions before and after the output modification; it does not imply agreement with the recorded ground-truth $u$.
The dynamic model in Fig.~\ref{fig:identification-perturbation} also has prediction errors with respect to the ground-truth $u$.

The models compared are selected checkpoints trained separately in Stage 2.
The batch size is 64 for both, while the learning rates differ: $3\times10^{-5}$ for dynamic and $10^{-5}$ for complete.
For dynamic, we used the version that also applies the graph constraints to the REG tokens; models after the implementation change that fully connects the REG tokens to one another are not included.
What this evaluation shows is the consistency between the known causal direction and the prediction dependencies in this identification example.
Because the weights and learning rates of the two models differ, the observed difference is not isolated as a causal effect of the graph mask alone.

Comparison conditions, raw arrays, and check results \(\cdot\) Vector version of Fig.~\ref{fig:identification-perturbation}

%% file: references.bib
@article{Ansari2024Chronos,
  author = {Ansari, Abdul Fatir and Stella, Lorenzo and Turkmen, Caner and Zhang, Xiyuan and Mercado, Pedro and Shen, Huibin and Shchur, Oleksandr and Rangapuram, Syama Sundar and Pineda Arango, Sebastian and Kapoor, Shubham and Zschiegner, Jasper and Maddix, Danielle C. and Wang, Hao and Mahoney, Michael W. and Torkkola, Kari and Wilson, Andrew Gordon and Bohlke-Schneider, Michael and Wang, Yuyang},
  title = {{Chronos}: Learning the Language of Time Series},
  journal = {Transactions on Machine Learning Research},
  year = {2024},
  issn = {2835-8856},
  eprint = {2403.07815},
  archivePrefix = {arXiv}
}

@techreport{Ansari2025Chronos2,
  author = {Ansari, Abdul Fatir and Shchur, Oleksandr and K{\"u}ken, Jaris and Auer, Andreas and Han, Boran and Mercado, Pedro and Rangapuram, Syama Sundar and Shen, Huibin and Stella, Lorenzo and Zhang, Xiyuan and Goswami, Mononito and Kapoor, Shubham and Maddix, Danielle C. and Guerron, Pablo and Hu, Tony and Yin, Junming and Erickson, Nick and Desai, Prateek Mutalik and Wang, Hao and Rangwala, Huzefa and Karypis, George and Wang, Yuyang and Bohlke-Schneider, Michael},
  title = {Chronos-2: From Univariate to Universal Forecasting},
  institution = {Amazon Web Services},
  year = {2025},
  eprint = {2510.15821},
  archivePrefix = {arXiv}
}

@article{Bittencourt2015NormalOperatingData,
  author = {Bittencourt, Andr{\'e} C. and Isaksson, Alf J. and Peretzki, Daniel and Forsman, Krister},
  title = {An Algorithm for Finding Process Identification Intervals from Normal Operating Data},
  journal = {Processes},
  volume = {3},
  number = {2},
  pages = {357--383},
  year = {2015},
}

@article{Busetto2024StateEstimators,
  title = {In-context learning of state estimators},
  author = {Busetto, Riccardo and Breschi, Valentina and Forgione, Marco and Piga, Dario and Formentin, Simone},
  journal = {IFAC-PapersOnLine},
  volume = {58},
  number = {15},
  pages = {145--150},
  year = {2024},
}

@article{Campi2002VRFT,
  author = {Campi, M. C. and Lecchini, A. and Savaresi, S. M.},
  title = {Virtual Reference Feedback Tuning: A Direct Method for the Design of Feedback Controllers},
  journal = {Automatica},
  volume = {38},
  number = {8},
  pages = {1337--1346},
  year = {2002},
}

@inproceedings{Carroll2022UniMASK,
  author = {Carroll, Micah and Paradise, Orr and Lin, Jessy and Georgescu, Raluca and Sun, Mingfei and Bignell, David and Milani, Stephanie and Hofmann, Katja and Hausknecht, Matthew and Dragan, Anca and Devlin, Sam},
  title = {{Uni[MASK]}: Unified Inference in Sequential Decision Problems},
  booktitle = {Advances in Neural Information Processing Systems},
  volume = {35},
  pages = {35365--35378},
  year = {2022},
}

@article{Dankers2016PredictorInputSelection,
  author = {Dankers, Arne and Van den Hof, Paul M. J. and Bombois, Xavier and Heuberger, Peter S. C.},
  title = {Identification of Dynamic Models in Complex Networks With Prediction Error Methods: Predictor Input Selection},
  journal = {IEEE Transactions on Automatic Control},
  volume = {61},
  number = {4},
  pages = {937--952},
  year = {2016},
}

@inproceedings{Das2024TimesFM,
  author = {Das, Abhimanyu and Kong, Weihao and Sen, Rajat and Zhou, Yichen},
  title = {A decoder-only foundation model for time-series forecasting},
  booktitle = {Proceedings of the 41st International Conference on Machine Learning},
  series = {Proceedings of Machine Learning Research},
  volume = {235},
  pages = {10148--10167},
  publisher = {PMLR},
  year = {2024},
  eprint = {2310.10688},
  archivePrefix = {arXiv}
}

@article{Forgione2023ICLSysId,
  author = {Forgione, Marco and Pura, Filippo and Piga, Dario},
  title = {From System Models to Class Models: An In-Context Learning Paradigm},
  journal = {IEEE Control Systems Letters},
  volume = {7},
  pages = {3513--3518},
  year = {2023},
}

@article{Forssell1999ClosedLoop,
  author = {Forssell, Urban and Ljung, Lennart},
  title = {Closed-loop Identification Revisited},
  journal = {Automatica},
  volume = {35},
  number = {7},
  pages = {1215--1241},
  year = {1999},
}

@article{Guo2019ASTGCN,
  author = {Guo, Shengnan and Lin, Youfang and Feng, Ning and Song, Chao and Wan, Huaiyu},
  title = {Attention Based Spatial-Temporal Graph Convolutional Networks for Traffic Flow Forecasting},
  journal = {Proceedings of the AAAI Conference on Artificial Intelligence},
  volume = {33},
  number = {01},
  pages = {922--929},
  year = {2019},
}

@misc{Jain2026TimesFM3,
  author = {Jain, Ayush and Sen, Rajat},
  title = {{TimesFM-3}: A zero-shot foundation model for multivariate forecasting},
  howpublished = {Google Research blog},
  year = {2026},
  month = aug,
  day = {31},
}

@article{Levine2020OfflineRL,
  author = {Levine, Sergey and Kumar, Aviral and Tucker, George and Fu, Justin},
  title = {Offline Reinforcement Learning: Tutorial, Review, and Perspectives on Open Problems},
  journal = {arXiv preprint arXiv:2005.01643},
  year = {2020},
  eprint = {2005.01643},
  archivePrefix = {arXiv},
  primaryClass = {cs.LG},
}

@article{Li2024OpenCity,
  author = {Li, Zhonghang and Xia, Long and Shi, Lei and Xu, Yong and Yin, Dawei and Huang, Chao},
  title = {Open Spatio-Temporal Foundation Models for Traffic Prediction},
  journal = {ACM Transactions on Intelligent Systems and Technology},
  year = {2025},
  month = oct,
  eid = {3773912},
  publisher = {Association for Computing Machinery},
}

@article{Markovsky2017MissingData,
  author = {Markovsky, Ivan},
  title = {A Missing Data Approach to Data-Driven Filtering and Control},
  journal = {IEEE Transactions on Automatic Control},
  volume = {62},
  number = {4},
  pages = {1972--1978},
  year = {2017},
}

@article{VanDenHof2013DynamicNetworks,
  author = {Van den Hof, Paul M. J. and Dankers, Arne and Heuberger, Peter S. C. and Bombois, Xavier},
  title = {Identification of Dynamic Models in Complex Networks with Prediction Error Methods---Basic Methods for Consistent Module Estimates},
  journal = {Automatica},
  volume = {49},
  number = {10},
  pages = {2994--3006},
  year = {2013},
}

@inproceedings{Wen2025M3PC,
  title = {{$M^3PC$}: Test-Time Model Predictive Control for Pretrained Masked Trajectory Model},
  author = {Wen, Kehan and Hu, Yutong and Mu, Yao and Ke, Lei},
  booktitle = {The Thirteenth International Conference on Learning Representations},
  year = {2025},
}

@inproceedings{Woo2024Moirai,
  author = {Woo, Gerald and Liu, Chenghao and Kumar, Akshat and Xiong, Caiming and Savarese, Silvio and Sahoo, Doyen},
  title = {Unified Training of Universal Time Series Forecasting Transformers},
  booktitle = {Proceedings of the 41st International Conference on Machine Learning},
  series = {Proceedings of Machine Learning Research},
  volume = {235},
  pages = {53140--53164},
  publisher = {PMLR},
  year = {2024},
  eprint = {2402.02592},
  archivePrefix = {arXiv}
}

@inproceedings{Wu2023MTM,
  author = {Wu, Philipp and Majumdar, Arjun and Stone, Kevin and Lin, Yixin and Mordatch, Igor and Abbeel, Pieter and Rajeswaran, Aravind},
  title = {Masked Trajectory Models for Prediction, Representation, and Control},
  booktitle = {Proceedings of the 40th International Conference on Machine Learning},
  series = {Proceedings of Machine Learning Research},
  volume = {202},
  pages = {37607--37623},
  year = {2023},
  publisher = {PMLR},
}

@inproceedings{Yu2018STGCN,
  author = {Yu, Bing and Yin, Haoteng and Zhu, Zhanxing},
  title = {Spatio-Temporal Graph Convolutional Networks: A Deep Learning Framework for Traffic Forecasting},
  booktitle = {Proceedings of the Twenty-Seventh International Joint Conference on Artificial Intelligence, {IJCAI-18}},
  publisher = {International Joint Conferences on Artificial Intelligence Organization},
  pages = {3634--3640},
  year = {2018},
}

@inproceedings{Yuan2025UrbanDiT,
  author = {Yuan, Yuan and Han, Chonghua and Ding, Jingtao and Zhang, Guozhen and Jin, Depeng and Li, Yong},
  title = {Diffusion Transformers as Open-World Spatiotemporal Foundation Models},
  booktitle = {Advances in Neural Information Processing Systems},
  volume = {38},
  year = {2025},
}

@inproceedings{Zhong2025FactoST,
  author = {Zhong, Siru and Qiu, Junjie and Wu, Yangyu and Zou, Xingchen and Rao, Zhongwen and Yang, Bin and Guo, Chenjuan and Xu, Hao and Liang, Yuxuan},
  title = {Learning to Factorize Spatio-Temporal Foundation Models},
  booktitle = {Advances in Neural Information Processing Systems},
  volume = {38},
  year = {2025},
}

@article{vanWaarde2020DataInformativity,
  author = {van Waarde, Henk J. and Eising, Jaap and Trentelman, Harry L. and Camlibel, M. Kanat},
  title = {Data Informativity: A New Perspective on Data-Driven Analysis and Control},
  journal = {IEEE Transactions on Automatic Control},
  volume = {65},
  number = {11},
  pages = {4753--4768},
  year = {2020},
}
